# Social Media and Health Misinformation during the US COVID Crisis
Centre for Democratic Engagement, University of Leeds, 20 July 2020

Gillian Bolsover
University of Leeds
g.bolsover@leeds.ac.uk

Janet Tokitsu Tizon
University of Leeds
janet.tokitsu@gmail.com

**Abstract**

*Health misinformation has been found to be prevalent on social media, particularly in new public health crises in which there is limited scientific information. However, social media can also play a role in limiting and refuting health misinformation. Using as a case study US President Donald Trump's controversial comments about the promise and power of UV light- and disinfectant-based treatments, this data memo examines how these comments were discussed and responded to on Twitter. We find that these comments fell into established politically partisan narratives and dominated discussion of both politics and COVID in the days following. Contestation of the comments was much more prevalent than support. Supporters attacked media coverage in line with existing Trump narratives. Contesters responded with humour and shared mainstream media coverage condemning the comments. These practices would have strengthened the original misinformation through repetition and done little to construct a successful refutation for those who might have believed them. This research adds much-needed knowledge to our understanding of the information environment surrounding COVID and demonstrates that, despite calls for the depoliticization of health information in this public health crisis, this is largely being approached as a political issue along divisive, polarised, partisan lines.*

**Health (mis)information during COVID**

Originating in Wuhan, China in December 2019, the spread of COVID-19 (a novel coronavirus) has created a global pandemic. The US has been the hardest hit country, with over 3 million confirmed cases and 135,000 deaths as of July 15th, 2020; the disease continues to spread rapidly in the US, with data suggesting the infections have yet to peak (CDC, 2020).

Due to the novelty of COVID-19, there is a significant lack of existing research on causes, effects, prevention and treatment. This has led to a circulation of health misinformation on social media as well as major news networks and government announcements. In March, a poll of 1,500 adult US citizens found that 13% of respondents believed the coronavirus was definitely or probably a hoax and 44% believe that the threat of the coronavirus was definitely or probably being exaggerated for political reasons (Economist/YouGov, 2020).

Social media posts at the time promoted these perspectives, calling the virus a "sham-demic" and alleging that it was being exaggerated by the Democratic Party for political gain. A March study found that more than a quarter of YouTube videos contained non-factual information about COVID (Oi-Yee et al., 2020). Another study found that Americans who used social media heavily were less likely to be able to correctly identify fake news articles as fake; Democrats were better able to correctly identify a news headline as true or false compared to Republicans and independents (Kreps & Kriner, 2020a).

A particularly notable case of health misinformation occurred when US President Donald Trump appeared to propose scientific research into treatments to expose the body to high amounts of UV light and to inject disinfectants in the body, which are known to be extremely dangerous (BBC, 2020). In a press briefing on the evening of April 23rd, 2020 Trump said, often speaking from the podium to Dr Deborah Birx, the White House coronavirus response co-ordinator:

*"So, supposing we hit the body with a tremendous, whether it's ultraviolet or just very powerful light, and I think you said that hasn't been checked but you're going to test it. And then I said, supposing you brought the light inside of the body, which you can do either through the skin or in some other way. And I think you said you're going to test that too. Sounds interesting. And then I see the disinfectant where it knocks it out in a minute. One minute. And is there a way we can do something like that, by injection inside or almost a cleaning? As you see it gets on the lungs, and it does a tremendous number on the lungs. So it'd be interesting to check that. So that, you are going to have to use medical doctors with, but it sounds interesting to me. So we'll see, but the whole concept of the light the way it kills it in one minute that's pretty powerful."*

Trump's statements about injecting disinfectant produced an immediate public response with the parent company of disinfectants Dettol and Lysol giving an official statement condemning the administration of disinfectant products into the human body (RB, 2020). After widespread concern and ridicule of these comments, Trump claimed that these comments were made sarcastically (Chiu et al., 2020).

Trump has also endorsed the use of the drug hydroxychloroquine for COVID-19 prevention, including announcing in May that he was taking the drug himself as a preventative measure and that it had been "determined (that the drug)… doesn't harm you"; this was, however, in contravention of FDA guidelines that stated that the drug should only be used in hospital settings due to the potential for serious side effects (McCarthy & Greve, 2020).

These comments are in keeping with a long history of inflammatory and misleading comments from the US President. Many of his statements have been declared false by third-party and major news network fact checkers such as politifact.com, factcheck.org and the Washington Post, CNN and NY Times. In the early stages of a global pandemic, it is important to remember that health misinformation can be highly detrimental and deadly where little information is known.

Although poisonings with bleach and other disinfectants had been on the rise in the US throughout 2020, having been touted as a medicine under the name "Miracle Mineral Supplement," poisonings appeared to rise even more dramatically after Trump's controversial comments (Kluger, 2020). Several US localities also reported significant rises in poison control centre calls (Glatter, 2020). Similarly, a 61-year-old Arizona man died and his 68-year-old wife was taken to hospital in critical condition after they both drank chloroquine phosphate (normally used to clean aquariums) after they learnt about the apparent disease-fighting properties of hydroxychloroquine a Trump press conference (Waldrop et al., 2020).

**Social media and health misinformation**

In academic research, social media has been shown to be a mechanism to circulate health misinformation (Wang et al., 2019). In health crises, this misinformation can lead to public frustration, confusion, and resistance to evidence-based health recommendations and science-based health information (Tan et al., 2015). In the case of the Ebola epidemic in West Africa in 2014, health misinformation circulated widely on social media, with Twitter feeds showing posts stating that Ebola could be cured with blood transfusions or by the subtropical plant ewedu (ibid).

Similarly, in the case of the Zika Virus in 2015, the lack of scientific research into the virus at the time led to false claims on social media that GM mosquitos brought Zika to Brazil and that government vaccines were responsible for birth defects (Bode & Vraga, 2018). These false claims play on existing scientifically refuted worries about genetically modified foods and the risks of vaccines. This is similar to the COVID case in which, even prior to Trump's 23 April comments, social media users had promoted drinking dangerously toxic chlorine dioxide solution as a cure/prevention (Mian & Kahn, 2020), repurposing the previously touted "Miracle Mineral Supplement" in this new context. Outside of public health crises such as Ebola, Zika and now COVID, health misinformation has long thrived on social media, for instance, driving a resurgence of the anti-vaccination movement (Smith & Graham, 2019).

Although health misinformation has been shown to be both prevalent and popular on social media

(Oi-Yee et al., 2020; Wang et al., 2019), social media can also play a role in countering misinformation. In the context of the Zika virus outbreak, it was found that corrective information on social media (both algorithmic and shared by other users) limited misperceptions among individuals exposed to that misinformation on social media (Bode & Vraga, 2018). In the context of COVID, a social media experiment found that simply labelling headlines as false had little effect on people's assessment of the accuracy of the information or their interactions with the information on social media (Kreps & Kriner, 2020b). Explicitly countering the misinformation proved more effective; however, some individuals continued to believe the false information after it had been explicitly countered (ibid).

In addition to the danger of believing and sharing false information, research has also suggested that viewing political misinformation can increase polarisation (DFR Lab, 2020) and that even when misinformation is corrected it does little to change the underlying attitudes and beliefs associated with the misinformation, suggesting these are driven by motivated reasoning (Hopkins et al., 2018; Swire et al., 2017). Another issue potentially underpinning the persistence of misinformation is that repetition in the context of a correction or retraction inadvertently makes the misinformation more familiar and thus strengthens its persistence (Ecker et al., 2017). However, not repeating misinformation when attempting to correct it, poses a particular problem for news reporting.

Journalistic principles have been blamed for helping Trump's presidential campaign, for instance, by reporting Trump's frequent explosive Twitter comments as news thereby providing large amounts of free coverage (Chadwick, 2017). Similarly, in the case of climate change, the journalistic principle of providing both sides of a story has been seen as keeping alive doubt about the existence and human causes of climate change long after overwhelming scientific consensus on these issues was established (Boykoff, 2004). How the news media as well as scientific and business opinion leaders phrase their reactions to the apparent proposition of dangerous and deadly treatments for the coronavirus by Trump is therefore crucial in the extent to which this misinformation might be countered or continued.

**Research approach**

With the world facing an unprecedented (if not un-predicted) global pandemic, there still remain many unknowns in how information about the disease is spreading on social media and how this information is being responded to by the general public, traditional media and opinion leaders. With success in fighting disease spread resting on individual-level actions and the majority of Americans getting news from social media (Gottfried & Shearer, 2016), the social media information environment will play a core role in whether individual-level actions, such as quarantining and mask-wearing, are successfully adopted.

Drawing from this provocation and to contribute to this important and fast-evolving topic area, this data memo will present data and analysis of US social media discourse surrounding Trump's comments on 23 April. Based on the reviewed literature, we focus on the following questions:

**1.** How prevalent was discourse about Trump's comments on social media? Did the majority of this discourse contest or promote these comments?
**2**. Did this misinformation appear to lead to public frustration, confusion and political polarisation?
**3.** How and to what extent did existing opinion leaders such as the traditional media, health and science professionals and business leaders respond to the spread of these comments on social media?
**4.** Did this response repeat and therefore could it inadvertently strengthen this misinformation through repetition and increased familiarity?
**5.** In public responses to the efforts of existing opinion leaders to counter this misinformation online, do we see evidence of public resistance to evidence-based and science-based health information?

**Methods and data collection**

In order to investigate these questions, we focus on discourse on Twitter. Although Twitter is not the largest social media platform in the US, it is the one most focused on news and current affairs. It is also the most open major social media platform and, therefore, an appropriate venue for studying a public discourse ecosystem rather than communications within delineated groups.

Many research projects intending to monitor social media discourse on Twitter collect data from within a set group of pre-selected hashtags and keywords. However, this strategy risks missing emergent or unselected topics. It is thus severely limited in its ability to speak to the body of online discourse, particularly during fast-moving events, and is subject to significant researcher bias based on the selection of hashtags and keywords to follow. To avoid this limitation, this project collected a sample of data from all trending topics within the US during the study period. The study period was chosen as 23 through 26 April to include the day on which Trump made these comments and the three following days, to take in media, individual and opinion leader reactions to these comments.

Using custom Python scripts to interface with the Twitter API, the project collected the most recent 100 tweets associated with each of the top 50 trending topics in the 64 locations for which Twitter collates trends (including one for the entire country) every 15 minutes during the target period: 23 through 26 April 2020. This data collection captured 2,041 unique trending topics across the four-day period.

We performed a content analysis of the 200 most popular of these trends to determine their topical content, based on the majority of tweets made within that trend during the time-period. The topical coding scheme used an established coding frame for social media posts developed by the researcher and used in previous studies in the US and China (Bolsover, 2017, 2018). Intercoder reliability tests were performed on a subset of these trends. Percentage agreement was 84% with a Kappa of 67%. These reliabilities are well within established appropriate ranges for this type of research (Lombard et al., 2002).

**Prevalence of COVID health information**

Of the 200 largest trends during this period, 38 concerned political topics (19%), 146 commercial topics (73%), 15 informational topics (7.5%) and 1 personal (private) topics (0.5%). Of the 38 political topics, 24 (63%) concerned COVID of which 17 (45%) contained COVID-related health information. Intercoder reliabilities for COVID and COVID-health information codes were percentage agreement 95% and 89% and Kappa 78% and 44%[1] respectively.

Taking a broad brush look at the information being shared on Twitter across this four-day period, we note that even during this time of extraordinary public health crisis, discussion of the disease only accounted for 12.5% of the top 200 trending topics[2]. Commercial trending topics, which includes products, services and entertainment, still accounted for almost three quarters of the top 200 trends. However, we note that the National (American) Football League (NFL) draft took place during this timeframe and received large amounts of discussion on Twitter, possibly explaining the large representation of commercial trends in the dataset. As such, health information, whether scientifically-backed or misinformation, was the major focus of less than 10% of the top 200 Twitter trends during this period. However, this still accounts for a large volume of information that Twitter users were exposed to and it is important to consider the content of these trends in order to address our research questions.

In order to understand the nature of this information, we randomly selected 250 posts from within the 38 political trends. We focus on all the political trends, rather than the subset about COVID or health information, because the content of trends is extremely varied and we are interested in including a wide-spread of political discourse, rather than simply the trends in which the majority of tweets constituted the selected code. Although 250 is a small sample of the more than 450,000 tweets across these 38 trends during the four-day period, this data memo is designed to present initial research into this important and fast-moving subject area. More in-depth research of larger samples and longer-time periods is, of course, needed and will be forthcoming. However, in a world of rapidly spreading misinformation on social media, it is important to make available initial reviews of datasets such as these to contribute to public discussion as it is happening and shape public, media and policy practice on this important issue.

**Content of discussion of Trump's comments**

Within these 250 tweets, there was a large amount of irrelevant information. This occurs for several reasons. Firstly, we collected the most recent 100 tweet that would show up in a search for the trending topic hashtag or keyword on Twitter. However, Twitter often will show unrelated popular or trending content alongside a search. This means that content about, for instance, the NFL draft, will show up when searching for COVID related topics on Twitter. Secondly, this also occurs because some keywords take in a variety of information. For instance, the trending keyword "stay" was coded as a political attempt to exert influence on individuals concerning COVID, as the majority of tweets focused on an exertion to "stay at home." However, the word can be associated with many non-political, non-COVID topics, such as staying in a relationship or staying somewhere on holiday. Lastly, irrelevant tweets also showed up in this dataset due to the practice of using trending hashtags and words unrelated to the content of the tweet to gain exposure.

Of the 250 tweets, 107 (43%) concerned US politics and 87 (35%) concerned COVID in the US. A number of posts concerned COVID in other countries, particularly in Latin America. These were excluded from consideration as this research focuses on the US case.

Posts that contested health misinformation and Trump's comments were more than twice as prevalent as posts that supported this health misinformation and Trump's comments, with 20 posts promoting health misinformation and 50 contesting[3]. With 70 out of 87 posts concerning COVID focusing on either supporting or contesting Trump's comments and related health misinformation, these comments dominated discussion about COVID on Twitter in the days following. These posts largely followed a partisan line, interpreting the comments as part of a wider media attack on President Trump. The most common was a retweet of a post that read:

*Trump used the word "disinfectant" meaning medicine that kills the virus. The media reports he said to drink bleach and huff Lysol. They are the enemy of the people. The enemy of the world.*

Another took a similar line, saying:

*Nancy Pelosi said this about @realDonaldTrump : "The president is asking people to inject Lysol into their lungs." This is an absolute lie. He never said that! Not even close. The liberal media won't call her out because they're too busy bashing Trump everyday. @POTUS*

Although the majority of the posts coded as supporting health misinformation were generic support of Trump in his comments and attacks on media reporting of the comments, some more general conspiracy theories were present in the sample, such as:

*4. The Coronavirus attack is exposed. The world knows it is an engineered virus from a lab in Wuhan, China. Democrats, Bill Gates, the WHO, China, and some folks at the CDC tried to fool us. The control is slipping, the Natives are restless. It's time for another BIG HEADLINE.*

or

*Bill Gates and the Depopulation Agenda: RFK Jr. Calls for an Investigation https://is.gd/qhVNxm #Coronavirus #Covid19 #CrimesAgainstHumanity #DeepState #KAG #MAGA #RFKJr #Stopbillgates #Vaccines #WWG1WGA*

Even a few outright conspiracy-theory posts in such a small sample (three out of 87 concerning COVID), suggests a worrying prevalence of these posts on social media.

In the 50 posts coded as contesting health misinformation, some took a partisan tone, with hashtags such as #TrumpIsALaughingStock. However, many more used humour to contest these comments, with posts such as:

*It turns out that Lysol kills self-serving, misleading press briefings 99.9 percent of the time!*

and

*At tonight's briefing trump will be introducing a new line of Trump Disinfectant in cherry lemon and orange flavor that MAGA can purchase for $49.99, if you purchase 3 or more you get a free trumpy Bear #TidePodPresident #disinfectant #Clorox #lysol*

Many posts also referred to a Saturday Night Live comedy skit in which actor Brad Pitt played leading member of the US coronavirus taskforce

Anthony Fauci despairing at having to so frequently correct Trump's misstatements about the virus. It is notable that instead of sharing words from Fauci or similar professionals, a humorous parody was shared frequently. Although much more entertaining and emotive, this parody repeated Trump's mis-informative claims about the virus and showed Fauci attempting to politely refute each of Trump's comments without actually contradicting the President. This tactic would not be likely to convince anyone who believed the initial statements and would contribute to the narrative of the liberal media attacking Trump. Although neither should be expected of a satirical comedy team, the fact that this is what was being shared rather than a scientific information or refutation is telling.

Despite this, a sizeable number of posts that contested the comments did share information from commercial and state voices to contest the comments. This included the retweeting of a CNN International post, a post by CNN anchor Jim Sciutto reporting Lysol's statement and an NBC New York report that New York City poison control calls doubled after Trump's comments. While posts supporting the comments were overwhelmingly personal opinions, a much larger percentage of posts that contested the comments shared news or information. Of posts contesting health misinformation, 20% shared political news or information, 72% political opinion and 8% political attempts to get individuals to act. In contrast, only 5% of posts supporting Trump's comments shared political news or information and 95% were political opinion.

Posts that contested the comments or other health misinformation were also much more likely to be pro-mainstream media rather than anti-mainstream media. Out of the 50 posts that contested, 10 (20%) displayed a pro-mainstream media orientation, while only 4 (8%) were anti-mainstream media. In contrast, of the 20 posts that supported the comments or other health misinformation only 1 (5%) was pro-mainstream media and 7 (35%) were anti-mainstream media. This is further evidence that responses to this information took partisan lines.

**Frustration, confusion and political polarisation?**

An analysis of the content of these posts suggest high levels of frustration and political polarisation, but very little confusion created by the comments. Rather than serving as a venue for the proliferation of misinformation per se, Twitter seems, in this case, to be more of a space in which existing partisan viewpoints can play out in this new case.

A significant proportion of posts concerning the comments were polarising and divisive. Of the 87 posts about COVID in the random sample of 250 posts in political trends, 91% were coded as attempting to prevent others from speaking or undermine the value of their words and 16% directed hate against a specific opposing group (that wasn't a political party).

This level of polarisation and hate is cause for concern in that the dissemination of health information surrounding a global pandemic should not be a political issue but rather one in which information and responses are de-politicised. However, in an election year, it is perhaps not surprising that contestation over the truth and interpretation of these comments has become highly polarised on Twitter.

The different sides in relation to these comments selectively employ different sources of evidence in line with ideas of the use of the Internet for ideologically driven history making through the assembling of facts and figures as part of a politically driven exercise (Udupa, 2016). Those who supported Trump's comments collated instances of media and politician extrapolation of Trump's comments, such as Lysol's announcement or Pelosi's comments that "The president is asking people to inject Lysol into their lungs," when he was suggesting, rather, research be conducted into the possible efficacy of these treatments about which he expressed excitement about their power and potential.

In contrast, those who contested Trump's comments drew sources from traditional media, who reported the voices of disinfectant manufacturers, poison control centres and medical professionals. However, as previously mentioned, contesting posts were much more likely to share news and information than supporting posts. For those that supported Trump's comments, the existence of a media, liberal, state and democratic bias against Trump was a fact, in and of itself. This draws from narratives Trump promotes of widespread conspiracies against him and his presidency (Trump, 2015), which are classic authoritarian techniques to generate emotionally based support and an easily recognisable enemy (Linz, 2000). The way traditional media responded to Trump's comments in this case could do little to alleviate this impression and also served to repeat the initial misinformation in a way that might serve to strengthen it.

**Repetition and response**

As discussed in the literature review attempts to counter misinformation may inadvertently end up strengthening its persistence through repetition (Ecker et al., 2017). With 80% of the posts about the coronavirus in the US and 61% of the posts about politics in the US either contesting or supporting Trump's comments or other related health misinformation, responses to these comments dominated online discourse in the days following.

None of the posts that contested the comments referred to methods of prevention or treatment of COVID for which there was scientific evidence at the time, such as social distancing, mouth covering, handwashing or treatments for which there had been some reported success in in-hospital scientific trials, such as the use of convalescent plasma to treat severely ill patients.

Rather posts reiterated the initial comments instructing individuals not to follow them and disparaging those who might believe the comments, such as this post from a medical doctor:

*I didn't think I would need to say this in 2020 but here we are. Do NOT inject yourselves with bleach. I repeat... DO NOT. #disinfectant #Covid_19 #coronavirus #COVID19 @realDonaldTrump*
*https://twitter.com/nytimes/status*

Traditional opinion leaders, such as CNN, NBC New York and Lysol, similarly repeated the initial claims in their refutations that were shared online. The headlines of shared news articles emphasised shock and simple refutation such as "Lysol maker: Please don't drink our cleaning products" (CNN) and "Please Don't Drink Disinfectant, Lysol And Dettol Maker Said After Trump Suggested People Could Inject It To Kill The Coronavirus" (Buzzfeed). The strength and incredulity of this reporting perhaps seems warranted given the danger of the suggested potential treatment and the fact that some people seemed prompted to experiment with the treatments Trump suggested scientists look into. However, the way it was responded to on social media both by users and the traditional media that they shared and retweeted would likely only increase the feelings of being attacked that underlie some of Trump's core election and re-election narratives.

Although there was no direct evidence of confusion or belief in these treatments in our Twitter sample, we should remember that these play into existing misinformation narratives in the US that touted bleach as a miracle cure prior to COVID and reports of increased bleach and disinfectant poison control instances after Trump's comments. The widespread coverage of the comments in the traditional media and the dominance of these comments in political and COVID-related social media discussion in the days following would have increased dissemination of the ideas to those who might be susceptible to believing them, or desperate or scared enough to try them.

**Developing resistance to evidence and science?**

As discussed in the literature review the circulation of misinformation can lead to resistance to evidence- and science-based information (Tan et al., 2015). Within the dataset, we largely do not see resistance to science- or evidence-based information. Largely discourse centred on reactions to the comments and then reactions to that reaction.

However, what was not present was, as previously mentioned, science or evidence-based alternative prevention and treatment techniques or the prevailing perspective of the scientific community that social distancing and enhanced hygiene measures would be necessary until more information and/or treatment options are available about the disease.

We also do not see much space given to scientific- or evidence-based voices in this discourse. Neither side needs science to fight their fight, with contesters arguing based on an assumption that what Trump appeared to suggest is ludicrous and supporters contesting the belief that Trump was genuinely suggesting scientific research into these potential treatments.

It is notable that while Anthony Fauci has been one of the most prominent scientific voices in the US, his voice was not present in the discourse but a humorous parody of Fauci's attempts to counter Trump's words about COVID (in which Fauci is played by actor Brad Pitt) was frequently mentioned. Similarly, another tweet that attempted to counter Trump's words, shared a video of White House coronavirus response co-ordinator Dr Deborah Birx, to whom Trump addressed and attributed some of his comments, as she watched Trump talk. The video appears to show her looking worried but saying nothing while Trump spoke.

The tendency to centre these serious discussions around politically polarised debate positions, ridicule, humour, entertainment and the words of politicians rather than scientists worryingly crowds out space for scientific- and evidence-based discourse. No direct evidence of resistance to science or evidence is found in the dataset. However, it is clear that these perspectives have little place in the polarised, emotive, pithy and rapidly-moving world of social media. Although previous research has suggested that social media can play a core role in refuting as well as disseminating health misinformation, there is no evidence that the kinds of information shown to work as refutation was being shared in this case.

**Limitations and further directions**

This data memo has presented an initial analysis of how a particular piece of health misinformation was discussed on US social media. With social media an increasingly important source of news and information for individuals across the world, the way information about COVID is discussed and shared on social media has important implications for the spread and persistence of the disease.

This research found that, while commercial topics made up more than three quarters of the trends during the study period, political and COVID discourse on Twitter was dominated by these comments. The way the comments were discussed fell into established patterns of political polarisation and served to repeat and amplify the initial comments. Posts showed worrying levels of attempting to shut down conversation and of directing hate at opposing groups. This case seems to suggest that COVID is being contextualised and discussed on social media as a largely political rather than scientific or public health issue.

This research is, of course, not without limitations. It is based on a single case of apparent health misinformation being disseminated by a political powerholder; a single social media platform (Twitter); and only a very small sample of the huge number of posts made in political trending topics during the period under consideration. The findings of this research should not be an end-point but can, perhaps, form a basis for broader and more in-depth research going forward.

However, we hope that the findings and analysis herein can form some temporary guidance and indications for those who might hope that social media could help disseminate information and stimulate positive discussions in the current pandemic. Although there was scant evidence of this being the case before COVID, many see the current crisis as an opportunity to change both civil society and politics for the better.

The major point of note here concerns how misinformation is approached, particularly when it is endorsed by powerful individuals, whether politicians, celebrities or other kinds of public individual. Analysis of Trump's 2016 campaign suggested that media coverage (even negative coverage) of his social media outbursts assisted his campaign and may even have been used as a concerted strategy to maintain a presence in the media (Wells et al., 2016). The importance of not affording greater space to inflammatory and mis-informative comments is even more crucial in a public health crisis, where misinformation can be a matter of life or death. In the current pandemic as well as more generally, we need to re-assess our belief that something is newsworthy because someone well-known says it.

Secondly, the level of divisiveness and polarisation surrounding these comments suggests a political orientation to the issue, driven by ideologically motivated reasoning and a politically driven selective use of facts to support ideological positions. There is little evidence that Twitter can provide the kind of space that would avert these deleterious processes. It is not the place to get news about a public health issue and discussing apparently newsworthy issues on the platform (although this might provide valuable entertainment and a feeling of connection in these socially distanced times) will ultimately do more harm than good.

Although there is some indication that readership of traditional news media has increased during COVID (WEF, 2020), it has also been found that 70% of US adults say they need to take breaks from COVID news (Mitchell et al., 2020). In this situation, it is crucial that public health information about the disease is disseminated in traditional formats, from non-political voices and in a scientifically-backed way. The discovery of new health information about the disease is not going to happen at a rate suitable for the 24-hour news media and social media environment. In the face of a very slow discovery of new information, the most responsible thing to do is to constantly reiterate existing scientifically-backed information to counter the misinformation that will inevitably arise as this process unfolds while refusing to engage with that misinformation and those voices that promote it.

---

[1] With COVID-related health information making up only a small proportion of the 200 trends, levels of expected agreement for health information are very high. This means that measures of intercoder reliability that take into account expected agreement, such as kappa, will be correspondingly lower.

[2] One trend concerned COVID but did not concern politics.

[3] Intercoder reliability tests have been performed on all reported quantitative data. These tests are deemed to fall within acceptable bounds for the rigour of test, complexity of coding frame and levels of expected agreement in the dataset. However, given the large number of categories reported here and the short length of this memo, intercoder reliability statistics for tweet codes are not reported. Several items, not reported here, were also assessed in the data but did not reach suitable levels of intercoder reliability and are, thus, not reported herein.

**Acknowledgements**

The authors wish to thank Emma Briones and Rhian Hughes for their content analysis work on the data; the University of Leeds Strategic Research Investment Fund for helping pay for student coding of project data; and the Laidlaw Scholarship program for funding and facilitating Tokitsu Tizon's work on the project.

Trump, D. (2015). *Crippled America: How to Make America Great Again*. Threshold Editions.

Udupa, S. (2016). Archiving as History-Making: Religious Politics of Social Media in India. *Communication, Culture & Critique*, *9*, 212–230.

Waldrop, T., Alsup, D., & McLaughlin, E. (2020, March 25). Fearing coronavirus, Arizona man dies after taking a form of chloroquine used to treat aquariums. *CNN*. https://edition.cnn.com/2020/03/23/health/arizona-coronavirus-chloroquine-death/index.html

Wang, Y., McKee, M., Torbica, A., & Stuckler, D. (2019). Systematic Literature Review on the Spread of Health-related Misinformation on Social Media. *Social Science and Medicine*, *240*(November 2019).

WEF. (2020). *The Media, Entertainment and Culture Industry's Response and Role in a Society in Crisis*. http://www3.weforum.org/docs/WEF_Media_Entertainment_Report_2020.pdf

Wells, C., Shah, D. V., Pevehouse, J. C., Yang, J., Pelled, A., Boehm, F., Lukito, J., Ghosh, S., & Schmidt, J. L. (2016). How Trump Drove Coverage to the Nomination: Hybrid Media Campaigning. *Political Communication*, *33*(4).